\documentclass[sigconf, screen]{acmart}
\usepackage{balance}
\usepackage{booktabs}
\usepackage{multirow}
\usepackage{amsmath,amsthm}
\usepackage{bm}
\usepackage{mathtools}
\usepackage{xcolor}
\usepackage{graphicx}
\usepackage{enumitem}
\usepackage{microtype}

\setcopyright{none}
\renewcommand\footnotetextcopyrightpermission[1]{}

\theoremstyle{definition}
\newtheorem{observation}{Observation}

\DeclareMathOperator{\Sinkhorn}{Sinkhorn}

\DeclareMathOperator{\reshape}{reshape}

\newcommand{\R}{\mathbb{R}}
\newcommand{\I}{\mathbf{I}}

\newcommand{\trans}{\mathsf T}
\newcommand{\Lrec}{\mathcal{L}_{\mathrm{rec}}}

\AtBeginDocument{%
  }

\begin{document}

\title{Stream-aware Side Adaptation for Large Pre-trained Multimodal Embedding Models in Sequential Recommendation}

\settopmatter{authorsperrow=3}

\author{Junchen Fu}
\affiliation{%
  \institution{University of Glasgow}
  \country{United Kingdom}
}
\email{j.fu.3@research.gla.ac.uk}

\author{Kaiwen Zheng}
\affiliation{%
  \institution{University of Glasgow}
  \country{United Kingdom}
}
\email{k.zheng.1@research.gla.ac.uk}

\author{Ioannis Arapakis}
\affiliation{
  \institution{Telef\'{o}nica Scientific Research, Telef\'{o}nica Innovaci\'{o}n Digital}
  \city{Barcelona}
  \country{Spain}
}
\email{arapakis.ioannis@gmail.com}

\author{Wenhao Deng}
\affiliation{%
  \institution{University of Glasgow}
  \country{United Kingdom}
}
\email{w.deng.1@research.gla.ac.uk}

\author{Xin Xin}
\affiliation{
\institution{Shandong University}\streetaddress{}\city{Qingdao}\country{China}}
\email{xinxin@sdu.edu.cn}

\author{Joemon M. Jose}\affiliation{
\institution{University of Glasgow}\streetaddress{}\city{Glasgow}\country{United Kingdom}}
\email{joemon.jose@glasgow.ac.uk}

\author{Xuri Ge}
\authornote{Corresponding Author.}
\affiliation{%
  \institution{Shandong University}
  \city{Jinan}\country{China}
}
\email{xuri.ge@sdu.edu.cn}

\renewcommand{\shortauthors}{Fu et al.}

\begin{abstract}
Recently, large pretrained multimodal embedding models such as Qwen3-VL Embedding have shown strong promise for sequential recommendation, as they provide reusable semantic item representations across modalities and domains. However, directly using these embeddings often leads to suboptimal performance because of domain misalignment. Efficient side adaptation is therefore an attractive solution. Although adapting all backbone layers should help, existing side adapters often degrade with depth, prompting layer dropping despite the loss of useful hidden states. This is due to two major challenges: (1) the lack of modeling in selecting fused representations during residual addition, and (2) the insufficient preservation of earlier representations during progressive sigmoid fusion. This paper therefore asks a practical question: \textit{How can we design a side adaptation approach that effectively unlocks the potential of large pre-trained multimodal embedding models?}

To address this question, we propose \emph{Stresa}, a stream-aware side-adaptation framework for frozen large pre-trained multimodal embedding models in sequential recommendation. Stresa introduces Stream-aware Hidden-Adapter Fusion (SHAF) to preserve historical side memory during fusion and Residual Stream Adapter (ReSA) to produce selective residual updates across layers. Empirically, Stresa consistently outperforms standard side adapters and state-of-the-art baselines on public datasets across multiple backbone embedding models. These results highlight the promise of adapting large embedding models for sequential recommendation. Our code is publicly available at~\url{https://github.com/GAIR-Lab/Stresa}.
\end{abstract}

\ccsdesc[500]{Information systems~Recommender systems}

\keywords{Multimodal Recommendation, Sequential Recommendation, Large Embedding Models, Side Adapter}

\maketitle

\section{Introduction}

\begin{figure}[t]
    \centering
    \includegraphics[width=\linewidth]{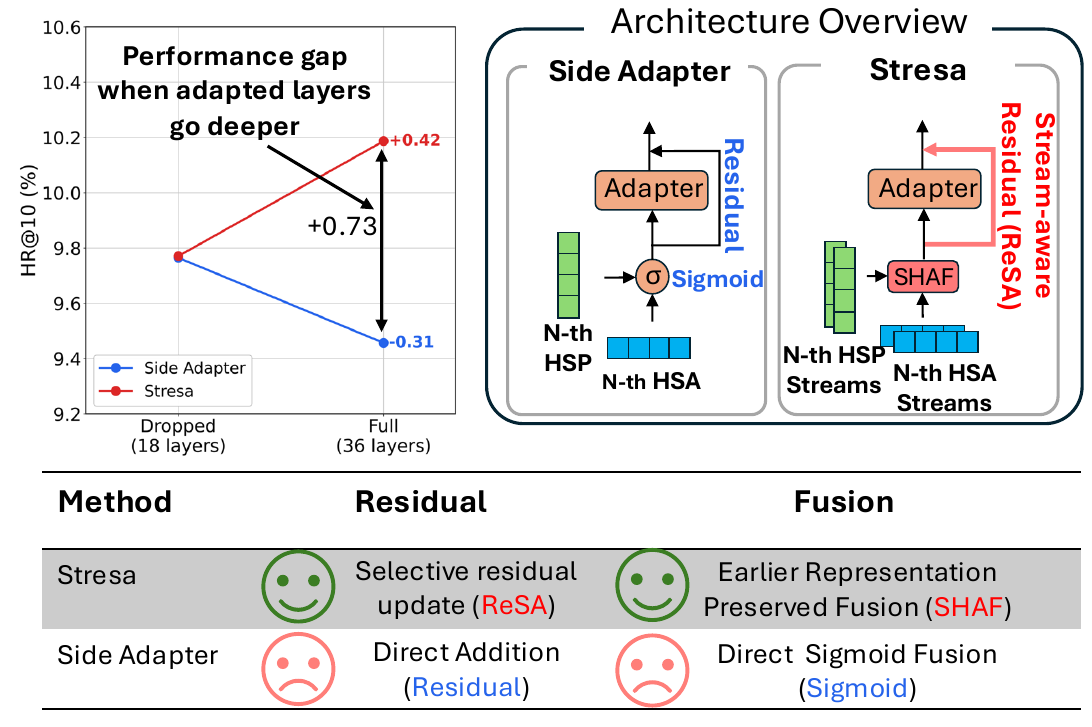}
    \caption{\textbf{Conventional Side Adapter vs.\ Stresa.} Using the frozen Qwen3-VL-8B-Embedding backbone, \emph{Dropped 18 layers} and \emph{Full 36 layers} mean that the item encoder extracts hidden states from 18 or 36 backbone layers for adapter computation. Left: HR@10 under both settings. Right: Side Adapter uses residual addition and sigmoid fusion, whereas Stresa uses selective residual update (ReSA) and earlier-representation-preserved fusion (SHAF). \textit{HSP} and \textit{HSA} denote hidden states from the pretrained model and previous adapter layer, respectively.}
    \label{fig:motivation}
\end{figure}

Large-scale pretrained multimodal embedding models are increasingly becoming reusable infrastructure for retrieval, ranking, and recommendation, as they project heterogeneous item content into transferable semantic spaces~\cite{radford2021learning,jia2021scaling,li2023blip,wang2022text,he2025double}. A recent state-of-the-art example is Qwen3-VL-Embedding, which demonstrates the power of this paradigm by mapping text, images, document images, and video into a unified representation space with strong multimodal retrieval performance~\cite{li2026qwen3}.
This is especially promising for multimodal sequential recommendation, where the quality of item's modality representations directly shapes next-item prediction~\cite{wang2025empowering,li2026capturing,zhou2025rethinking,hu2024bivrec,wang2025leveraging}. When the predictor operates on semantically rich multimodal embeddings, it can exploit content continuity, semantic substitutability, and cross-modal expressions of user intent in addition to collaborative signals~\cite{ni2025content,fu2024iisan,fu2025efficient,fu2025crossan}. Stronger multimodal embedding backbones should therefore provide a stronger foundation for sequential recommendation.

However, directly applying large pre-trained multimodal embedding models to sequential recommendation often leads to suboptimal performance due to domain misalignment~\cite{fu2024iisan}. These pretrained embedding models are typically learned for general semantic representation rather than recommendation-specific objectives, while sequential recommendation depends on user behavior, transition patterns, and recommendation relevance that can differ substantially from standard semantic similarity~\cite{yuan2023go}. Although end-to-end adaptation can effectively bridge this gap, it is often impractical for deployment because of its high computational and memory cost~\cite{li2025exploring}. This trade-off is especially important in sequential recommendation, where the same item encoder is reused over a large catalog and efficient offline embedding extraction is often crucial for deployment. This challenge remains significant not only for full fine-tuning but also for conventional parameter-efficient tuning when the backbone model is large~\cite{sung2022lst}. In this setting, side adaptation~\cite{fu2024iisan,fu2025crossan} is particularly appealing because it keeps the backbone frozen while learning only a lightweight side branch, enabling efficient training, offline caching of item representations, and practical deployment at scale, while still retaining strong recommendation performance. Motivated by this promise, recent works have begun to explore side-adapter-style designs for sequential recommendation with large embedding backbones~\cite{fu2024iisan,fu2025efficient,fu2025crossan,xu2023side}. In this work, we focus specifically on frozen, offline-cacheable item-side adaptation of a \textit{unified single-tower} multimodal embedding backbone. The resulting item representations are user-independent, while the downstream sequential model is treated as a shared component rather than part of the proposed adaptation architecture.

Most prior methods~\cite{fu2024iisan,fu2025efficient,fu2025crossan} adopt a conventional side-adapter design, in which a lightweight bottleneck adapter with a residual connection is attached to each backbone layer to adapt its hidden states. Although this paradigm is computationally efficient, we find that its performance becomes suboptimal when extended to full-depth adaptation, as shown in Figure~\ref{fig:motivation}. Existing studies~\cite{sung2022lst,fu2024iisan,fu2025efficient} observe the same issue and therefore treat layer dropping as a default setting, since adapting more hidden layers can even degrade performance.

 We argue that this limitation stems from two key challenges: (1) the lack of modeling of how fused representations are selected during direct residual addition, and (2) the inadequate preservation of earlier representations during progressive sigmoid fusion. \textbf{First}, residual connection lacks modeling of what to retain or suppress in fused representations. In side adaptation, the propagated representation at each layer is already a fused embedding of the current hidden states and the previous adapter's output~\cite{fu2024iisan,sung2022lst}. 
 For such representations, effective propagation requires deciding which components should be preserved and which should be suppressed. Plain residual addition does not model this process, which may limit the quality of the propagated embedding. \textbf{Second}, simple fusion~\cite{fu2024iisan,sung2022lst}
 cannot effectively preserve earlier representations during progressive aggregation. Since side adaptation proceeds by continuously incorporating newly emerging layer-wise information, the model should also retain useful representations accumulated from earlier layers. However, simple sigmoid-based fusion is often insufficient for this purpose, causing earlier information to be gradually overwritten as depth increases. These observations motivate the central question of this paper: \emph{how can we design effective side adaptation for deep unified multimodal embedding models?}

To this end, we propose \emph{Stresa}, \textbf{Stre}am-aware \textbf{S}ide \textbf{A}daptation for large pre-trained multimodal embedding models in sequential recommendation. Stresa addresses the two challenges above by redesigning the two core components of conventional side adapters: residual connection and sigmoid fusion. To enable modeling in residual connection, inspired by the empirical success of stream-aware residual designs in large language models~\cite{zhu2024hyper}, we introduce \emph{ReSA}, a residual stream adapter tailored for side-adapter blocks. To move beyond simple sigmoid fusion toward stronger fusion and better preservation of earlier-layer representations, we further propose \emph{Stream-aware Hidden-Adapter Fusion} (SHAF), which not only produces a stream-structured fused representation for ReSA, but also preserves historical side memory from earlier adapter layers. Together, ReSA and SHAF enable stronger fusion, more stable memory propagation, and more effective deep side adaptation.

Our contributions are summarized as follows:
\begin{itemize}[leftmargin=*]
    \item We propose \emph{Stresa}, a \textbf{stre}am-aware \textbf{s}ide \textbf{a}daptation paradigm that enables deep adaptation for unified large pre-trained multimodal embedding models.
    
    \item Stresa introduces the \emph{Residual Stream Adapter} (ReSA) for selection modeling in residual connection via stream-aware residual adaptation, and the \emph{Stream-aware Hidden-Adapter Fusion} (SHAF) for fine-grained stream-aware fusion with better preservation of earlier-layer representations.
    
    \item We provide both structural analysis and empirical evidence showing that Stresa better preserves earlier representations and remains effective under deep adaptation. Across multiple datasets and two Qwen3-VL-Embedding scales, Stresa consistently outperforms the standard Side Adapter and strong prior baselines.
\end{itemize}

\begin{figure*}[t]
    \centering
    \includegraphics[width=\textwidth]{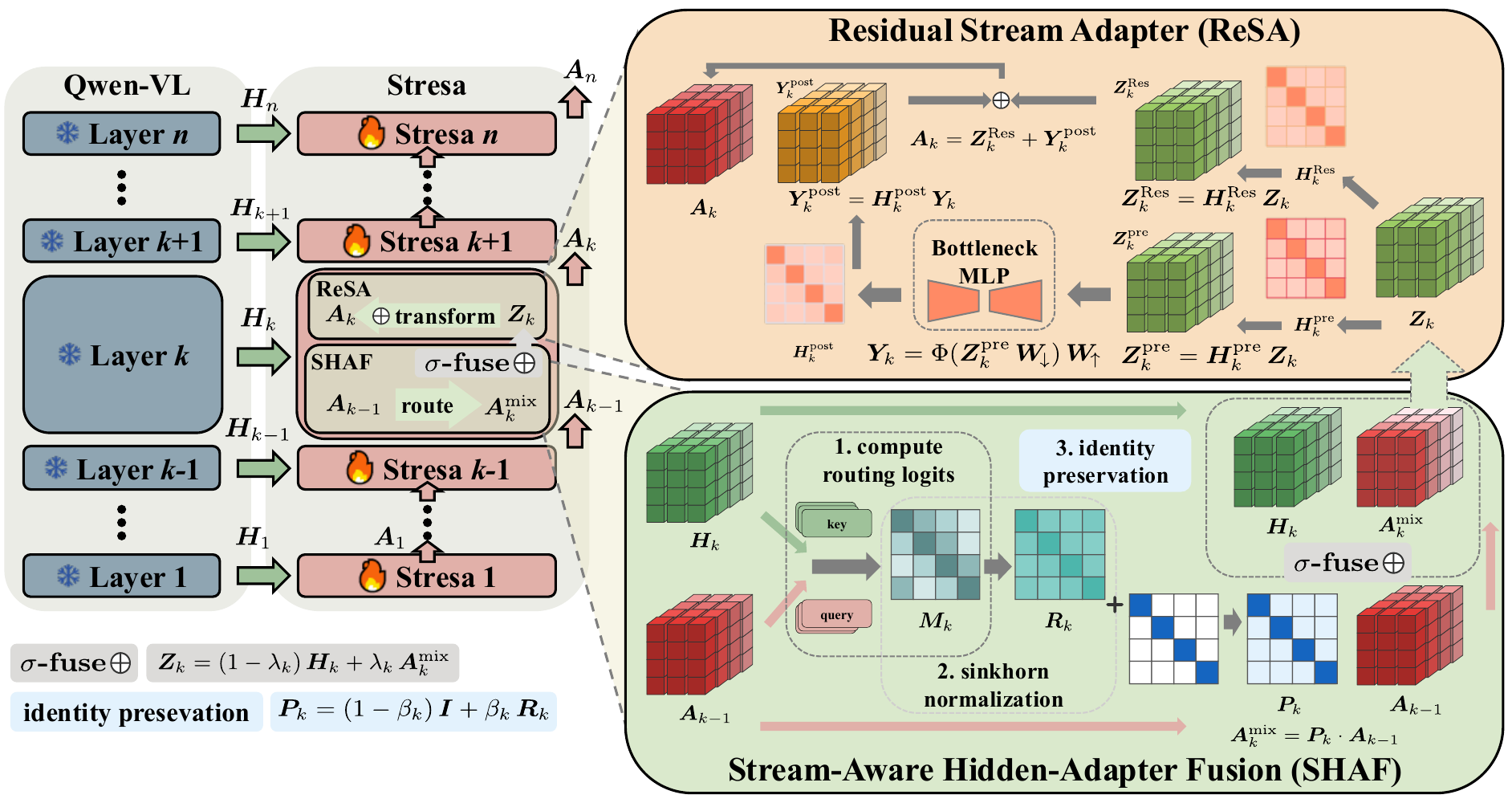}
    \caption{
Overview of the proposed stream-aware fusion framework.
(\textbf{1}) \textbf{Stream-aware Hidden-Adapter Fusion (SHAF)} at layer $k$ first reshapes the hidden representations into multiple streams and computes routing logits via query-key interactions. An identity-preservation operation is applied to preserve information from earlier layers.
(\textbf{2}) \textbf{Residual Stream Adapter (ReSA)} applies stream-wise pre-, post-, and residual mappings around a bottleneck MLP. The mapped residual and nonlinear branches are combined to produce the next side state $\mathbf{A}_k$, which is directly propagated to the following adapted layer.
}
    \label{fig:overview}
\end{figure*}
\section{Related Work}

\noindent \textbf{Pre-trained Multimodal Embedding Models.} Pretrained encoders first established the value of reusable contextual representations for downstream matching and ranking~\cite{devlin2019bert,liu2019roberta}. Multimodal representation learning then extended this idea to image--text and vision--language settings through models such as CLIP, ALIGN, BLIP-style architectures, and more recent general-purpose embedding backbones~\cite{radford2021learning,jia2021scaling,li2023blip,wang2022text,li2026qwen3,he2025double}. These models are increasingly used as reusable embedders, which makes them a natural foundation for recommendation when items contain rich text and visual content. Recently, the Qwen team proposed the Qwen3-VL-Embedding models~\cite{li2026qwen3}, which achieve state-of-the-art results in multimodal representation learning. This trend is particularly relevant to sequential recommendation. Recent multimodal recommendation benchmarks such as MicroLens~\cite{ni2025content} highlight that user behavior often depends on content-rich items, making large pre-trained multimodal embedding models a strong starting point for item representation. Recent recommendation-oriented adaptation frameworks further show that frozen or lightly adapted multimodal backbones can materially improve recommendation~\cite{liu2024alignrec,geng2023vip5,liu2024multimodal,kim2025lost,wang2025rethinking,bao2023tallrec,fu2024exploring,ye2026multimodal,zhuang2025frequency}. These embeddings have also been adopted in the generative retrieval paradigm for recommendation~\cite{sun2026zerogr,fu2026differentiable}.

\noindent \textbf{Parameter-Efficient Adaptation and Side Adaptation} Parameter-efficient adaptation methods such as adapters, prompt tuning, and low-rank updates reduce trainable parameter counts relative to end-to-end tuning~\cite{houlsby2019parameter,lester2021power,hu2022lora}. In recommendation, however, the relevant design criterion is not only parameter count. Because the same item encoder is reused across a large catalog, adaptation methods that preserve a frozen backbone are especially attractive for caching, deployment, and offline feature extraction. Side adaptation~\cite{sung2022lst,xu2023side} is a natural fit for this constraint. It learns a lightweight side adapter while keeping the backbone frozen, thereby avoiding the need to store the backbone’s backward computation graph. Subsequent work, including Quantized Side Tuning, further explored lightweight side branches as a way to complement frozen LLMs~\cite{zhang2024quantized}. In multimodal sequential recommendation, recent methods have shown that multimodal information can be used as cached embeddings to improve efficiency. For example, IISAN~\cite{fu2024iisan} introduces this paradigm, IISAN-Versa~\cite{fu2025efficient} extends it by adapting larger asymmetric two-tower encoders of different scales, and CROSSAN~\cite{fu2025crossan} further demonstrates the effectiveness of side adaptation across four modalities. These studies mainly focus on efficiency and adaptation design in multi-tower architectures. However, most of them rely on LayerDrop~\cite{sung2022lst} and do not address the stability of deep side-memory propagation in such models. Besides these, many state-of-the-art studies~\cite{zhang2024multimodal,ye2025harnessing,liang2023mmmlp,zhang2025hierarchical,song2025boosting,dai2026cross,lu2025dmmd4sr,meng2025tamer,yang2025fitmm,guo2025mmhcl,cui2025multi,guo2025m,zhu2025beyond,xu2025mentor} propose architectures based on multimodal large language models or related techniques to tackle multimodal recommendation tasks, but they do not investigate the side adaptation problem.

To the best of our knowledge, efficient and stable deep side adaptation for large pre-trained multimodal embedding models in sequential recommendation remains largely underexplored, which motivates the present study.

\section{Method}

\subsection{Problem Formulation}

Let $\mathcal{U}$ and $\mathcal{V}$ denote the user and item sets. For each user $u\in\mathcal{U}$, we observe a chronological interaction sequence
\begin{equation}
\mathcal{S}_u = (v_1^u, v_2^u, \ldots, v_{T_u}^u), \qquad v_t^u \in \mathcal{V}.
\end{equation}
Each item $v$ is associated with multimodal content
\begin{equation}
\mathbf{x}_v = \{\mathbf{x}_v^{(m)}\}_{m=1}^{M},
\end{equation}
where $M$ is the number of modalities. Let $\mathcal{B}$ be a frozen large multimodal embedding model with $L$ layers. Given $\mathbf{x}_v$, the backbone produces hidden states
\begin{equation}
\mathbf{h}_v^{(1)}, \mathbf{h}_v^{(2)}, \ldots, \mathbf{h}_v^{(L)}, \qquad \mathbf{h}_v^{(\ell)} \in \R^d.
\end{equation}
We choose a set of adapted layers
\begin{equation}
\mathcal{L}=\{\ell_1 < \ell_2 < \cdots < \ell_K\} \subseteq \{1,\ldots,L\}.
\end{equation}

Because the backbone is frozen, the selected hidden states can be extracted and cached offline. Stresa recursively constructs a task-specific side representation from these cached representations:
\begin{equation}
\mathbf{a}_v^{(0)}=\mathbf{0}, \qquad
\mathbf{a}_v^{(k)}
=
\mathrm{Stresa}_k
\left(
\mathbf{h}_v^{(\ell_k)},
\mathbf{a}_v^{(k-1)}
\right),
\quad k=1,\ldots,K,
\label{eq:memory_update}
\end{equation}
and produces the final adapted embedding
\begin{equation}
\tilde{\mathbf{e}}_v=\mathbf{e}_v^0+\mathbf{W}_o\mathbf{a}_v^{(K)},
\label{eq:final_embedding}
\end{equation}
where $\mathbf{e}_v^0\in\R^d$ is the frozen backbone output used by the downstream recommender.

A sequential recommender $\mathcal{F}_{\psi}$ maps the adapted history $(\tilde{\mathbf{e}}_{v_1^u},\ldots,\tilde{\mathbf{e}}_{v_t^u})$ to a user state $\mathbf{q}_{u,t}$, and the score for candidate item $v$ is
\begin{equation}
s(u,t,v)=\mathbf{q}_{u,t}^{\trans}\tilde{\mathbf{e}}_v.
\end{equation}
We optimize the standard next-item prediction loss
\begin{equation}
\Lrec = -\sum_{(u,t)} \log \frac{\exp(s(u,t,v_{t+1}^u))}{\sum_{v'\in\mathcal{C}_{u,t}} \exp(s(u,t,v'))},
\end{equation}
where only the Stresa parameters, the output projection $\mathbf{W}_o$, and the sequence model are trainable. Below, we omit the item index and write $\mathbf{h}_k:=\mathbf{h}^{(\ell_k)}$ and $\mathbf{a}_{k-1}:=\mathbf{a}^{(k-1)}$ for a single Stresa block.

\subsection{Overview of Stresa}

A standard side adapter typically fuses the current backbone state with the previous side state and then applies a lightweight transform. This works well for shallow adaptation, but under deeper adaptation the side branch is no longer just a temporary local feature: it becomes a memory that is propagated across layers. Once this happens, two questions become central: how should historical side memory arrive at the current layer, and how should the fused signal be transformed into the next side state while retaining useful historical information? Stresa addresses these two questions explicitly.

At adapted layer $k$, Stresa first aligns the propagated side memory with the current backbone representation through a stream-aware fusion module, and then converts the fused state into the next propagated side representation:
\begin{equation}
\mathbf{z}_k
=
\mathrm{SHAF}
\left(
\mathbf{h}_k,
\mathbf{a}_{k-1}
\right),
\qquad
\mathbf{a}_k
=
\mathrm{ReSA}
\left(
\mathbf{z}_k
\right).
\label{eq:Stresa_overall}
\end{equation}
Figure~\ref{fig:overview} illustrates the pipeline. SHAF transports the previous side representation to the current depth and fuses it with the current backbone state. ReSA then applies an HC-style residual composition to the fused representation and produces the next side state, which is directly propagated to the following adapted layer. In this sense, Stresa separates \emph{memory transport} from \emph{state refinement}: SHAF determines how the previous side state enters the current layer, while ReSA determines how the fused representation is transformed into the next propagated state.

This separation is the main design principle behind Stresa. The frozen backbone state $\mathbf{h}_k$ serves as the layer-local semantic anchor at depth $k$, whereas the side branch carries task-specific information propagated from earlier adapted layers. Stresa therefore treats deep adaptation as \emph{progressive state refinement}: each adapted layer first transports and aligns the previous side state with the current frozen representation, and then maps the fused signal into the next side state.

\subsection{Stream-aware Hidden-Adapter Fusion  (SHAF)}

SHAF is designed to address two needs in deep side adaptation: (1) stronger fusion with the current hidden state and (2) preservation of historical memory. Plain sigmoid fusion only controls how much previous side memory enters the current layer, but does not model how it should interact with the current hidden representation in a structured way. SHAF therefore introduces stream-aware interaction before fusion, so that the propagated memory can first be aligned with the current hidden state at the stream level before entering the final fusion. This makes the fusion more selective and informative, and also organizes the fused representation for the subsequent adapter. At the same time, since the side state is propagated across layers, it should remain an explicit memory path during deep adaptation. SHAF thus adopts an identity-preserving routing operator, which retains a direct carry-over route from the historical memory while allowing controlled cross-stream reorganization when useful.

We first reshape both signals into $s$ streams of width $d_s$, where $d=s\,d_s$:
\begin{equation}
\mathbf{H}_k=\reshape(\mathbf{h}_k)\in\R^{s\times d_s}, \qquad
\mathbf{A}_{k-1}=\reshape(\mathbf{a}_{k-1})\in\R^{s\times d_s}.
\label{eq:stream_reshape}
\end{equation}
This stream decomposition is a lightweight structured coordinate system rather than a semantic assumption about the hidden dimension. It allows SHAF to model cross-subspace interaction through small stream-level operators, and it also gives the following ReSA block the same structured space in which to refine the fused representation.

To decide how historical side memory should interact with the current layer, SHAF computes stream-level routing logits from pooled descriptors of $\mathbf{H}_k$ and $\mathbf{A}_{k-1}$:
\begin{equation}
\mathbf{M}_k(i,j)=\frac{\mathbf{q}_{k,i}^{\trans}\mathbf{k}_{k,j}}{\sqrt{d_r}},
\label{eq:routing_logits}
\end{equation}
where $\mathbf{q}_{k,i}$ and $\mathbf{k}_{k,j}$ are learned query/key projections of the corresponding stream summaries. We then apply Sinkhorn normalization~\cite{adams2011ranking},
\begin{equation}
\mathbf{R}_k = \Sinkhorn\!\left(\frac{\mathbf{M}_k}{\tau}\right),
\label{eq:sinkhorn_main}
\end{equation}
to obtain a balanced soft routing matrix over streams. This keeps the stream redistribution structured rather than leaving it as an unconstrained dense remapping.

Routing alone, however, would directly replace each stream by a mixture of other streams. For deep propagation, we instead use an identity-preserving routing operator:
\begin{equation}
\beta_k = \sigma(\hat{\beta}_k), \qquad
\mathbf{P}_k = (1-\beta_k)\I + \beta_k \mathbf{R}_k .
\label{eq:pk}
\end{equation}
This is the central design choice of SHAF. The gate $\beta_k$ controls how far the transported memory is allowed to deviate from its original stream layout: when $\beta_k$ is small, the side memory is carried forward mostly along its current stream path; when $\beta_k$ is larger, related streams can exchange information through $\mathbf{R}_k$. In other words, SHAF treats cross-stream routing as a controlled reorganization of historical memory rather than a hard replacement.

Finally, SHAF routes the historical side memory and fuses it with the current backbone state:
\begin{equation}
\mathbf{A}_k^{\mathrm{mix}} = \mathbf{P}_k \mathbf{A}_{k-1},
\end{equation}
\begin{equation}
\lambda_k = \sigma(\hat{\lambda}_k), \qquad
\mathbf{Z}_k = (1-\lambda_k)\mathbf{H}_k + \lambda_k \mathbf{A}_k^{\mathrm{mix}},
\label{eq:zk}
\end{equation}
where $\lambda_k$ controls the overall contribution of historical side memory relative to the current backbone state. The two gates $\beta_k$ and $\lambda_k$ therefore play different roles: $\beta_k$ controls \emph{how historical memory is redistributed across streams}, whereas $\lambda_k$ controls \emph{how strongly the transported memory enters the current layer}. This separation makes SHAF easier to interpret and also underlies the structural observations in Sec.~\ref{sec:theory}: SHAF can be viewed as plain sigmoid fusion together with a controlled stream-aware correction.

\subsection{Residual Stream Adapter (ReSA)}

After SHAF, the fused representation $\mathbf{Z}_k$ already lives in stream space. ReSA is designed to refine this structured signal directly, rather than collapsing it back to a single vector and applying a standard bottleneck update. Inspired by Hyper-Connections, ReSA introduces lightweight stream interaction before, after, and along the residual branch of the bottleneck transform.

The adapter first performs \emph{pre-connection mixing},
\begin{equation}
\mathbf{Z}_k^{\mathrm{pre}} = \mathbf{H}_k^{\mathrm{pre}}\mathbf{Z}_k,
\end{equation}
which lets each stream be transformed in the context of other related streams before entering the bottleneck. It then applies a shared bottleneck MLP,
\begin{equation}
\mathbf{Y}_k=\phi\!\left(\mathbf{Z}_k^{\mathrm{pre}}\mathbf{W}_{\downarrow}\right)\mathbf{W}_{\uparrow},
\label{eq:bottleneck}
\end{equation}
followed by \emph{post-connection mixing},
\begin{equation}
\mathbf{Y}_k^{\mathrm{post}} = \mathbf{H}_k^{\mathrm{post}}\mathbf{Y}_k,
\end{equation}
which redistributes the transformed features back across streams. Here $\mathbf{H}_k^{\mathrm{pre}}$ and $\mathbf{H}_k^{\mathrm{post}}$ are learned stream-mixing operators internal to ReSA; they are distinct from the reshaped backbone state $\mathbf{H}_k$ in Eq.~\eqref{eq:stream_reshape}.

Stresa also adds \emph{adaptive residual routing}:
\begin{equation}
\mathbf{Z}_k^{\mathrm{res}} = \mathbf{H}_k^{\mathrm{res}}\mathbf{Z}_k.
\label{eq:residual_routing}
\end{equation}
This residual route provides a direct stream-wise shortcut for the fused representation. It complements the nonlinear bottleneck path: the bottleneck increases expressive power, while the residual route preserves a simple path for information that does not need strong transformation.

The final ReSA output is
\begin{equation}
\mathbf{A}_k
=
\mathbf{Z}_k^{\mathrm{res}}
+
\mathbf{Y}_k^{\mathrm{post}} .
\label{eq:resa_output}
\end{equation}
Here $\mathbf{A}_k$ denotes the next stream-space side state produced by ReSA. It is directly propagated to the following adapted layer and serves as the historical side state in the next SHAF block. The residual operation is internal to ReSA: $\mathbf{Z}_k^{\mathrm{res}}$ provides the mapped shortcut branch, while $\mathbf{Y}_k^{\mathrm{post}}$ provides the nonlinear refinement branch.

Relative to a standard side adapter, ReSA provides two advantages. First, stream interaction occurs before, after, and along the residual branch of the bottleneck, making the refinement more structured than a single transform on an unstructured vector. Second, the fixed identity shortcut of a conventional residual adapter is replaced by a learnable stream-wise residual mapping. Overall, SHAF and ReSA play complementary roles: SHAF determines how the previous side state is transported and fused at the current depth, while ReSA maps the fused representation into the next propagated side state. These complementary roles are further clarified by the structural observations in Sec.~\ref{sec:theory}.

\subsection{Complexity and Deployment}

Compared with a plain side adapter, Stresa introduces additional computation only through lightweight stream-level operators. In SHAF, the extra cost comes from computing stream-level routing logits, applying Sinkhorn normalization, and multiplying the routed memory by the $s\times s$ operator $\mathbf{P}_k$. In ReSA, the additional structure comes from the stream-mixing operators $\mathbf{H}_k^{\mathrm{pre}}$, $\mathbf{H}_k^{\mathrm{post}}$, and $\mathbf{H}_k^{\mathrm{res}}$, together with the shared bottleneck MLP.

Per adapted layer, the stream operators act on matrices of size $s\times s$ or $s\times d_s$, while the bottleneck operates on width $d_s$ with rank $r$. Thus, the added overhead scales with the number of streams and the stream width, rather than with a dense full-dimensional $d\times d$ transform. In practice, this overhead is modest because typically $s \ll d$ and $r \ll d_s$.

This additional cost buys two capabilities that a plain side adapter does not model explicitly. SHAF introduces controlled transport of historical side memory before fusion, instead of injecting $\mathbf{a}_{k-1}$ as a single unstructured vector. ReSA then refines the fused representation in the same stream geometry through an explicit stream-wise residual composition within each Stresa block. The gain is therefore not just extra nonlinearity, but a more structured treatment of deep side adaptation.

Stresa also preserves the deployment pattern that matters in recommendation. Since the backbone is frozen, selected hidden states can be extracted and cached offline. Training backpropagates only through the lightweight side branch, the output projection $\mathbf{W}_o$, and the downstream sequential model. After training, the adapted item embeddings $\tilde{\mathbf e}_v$ remain item-specific and can be refreshed offline and served in the same way as standard item embeddings. In this sense, Stresa changes the adaptation mechanism without changing the frozen-backbone workflow preferred by large-scale recommendation systems. 

\subsection{Structural Observations on SHAF and ReSA}
\label{sec:theory}

We provide three algebraic observations that clarify the roles of SHAF and ReSA. They characterize how SHAF differs from plain sigmoid fusion and how ReSA relates to a conventional residual bottleneck adapter. These observations describe the forward mapping of Stresa and are not optimization or convergence guarantees.

\begin{observation}[SHAF as a correction to plain fusion]
\label{obs:shaf_extension}
Let
\(
\mathbf{Z}_k^{\mathrm{plain}}
=
(1-\lambda_k)\mathbf{H}_k
+
\lambda_k\mathbf{A}_{k-1}
\)
denote plain sigmoid fusion. The SHAF representation can be written exactly as
\begin{equation}
\mathbf{Z}_k
=
\mathbf{Z}_k^{\mathrm{plain}}
+
\mathbf{C}_k,
\qquad
\mathbf{C}_k
=
\lambda_k\beta_k
(\mathbf{R}_k-\mathbf{I})
\mathbf{A}_{k-1}.
\label{eq:theory_z_decomp_main}
\end{equation}
\end{observation}

Observation~\ref{obs:shaf_extension} shows that SHAF contains plain sigmoid fusion as the special case \(\beta_k=0\) or \(\mathbf{R}_k=\mathbf{I}\), while otherwise adding a controlled stream-aware correction.

\begin{observation}[Common-mode invariance of the SHAF correction]
\label{obs:shaf_common_mode}
Let
\(
\Pi=\frac{1}{s}\mathbf{1}\mathbf{1}^{\trans}
\)
be the projector onto the common stream mode. If \(\mathbf{R}_k\) is doubly stochastic, then
\begin{equation}
\mathbf{C}_k
=
\lambda_k\beta_k
(\mathbf{R}_k-\mathbf{I})
(\mathbf{I}-\Pi)
\mathbf{A}_{k-1},
\qquad
\Pi\mathbf{C}_k=\mathbf{0}.
\label{eq:theory_common_mode_main}
\end{equation}
\end{observation}

Thus, relative to plain sigmoid fusion, the additional SHAF routing correction leaves the common stream mode unchanged and acts only on variation across streams. With a finite number of Sinkhorn iterations, this statement holds approximately to the extent that \(\mathbf{R}_k\) is approximately doubly stochastic.

\begin{observation}[ReSA generalizes a residual bottleneck update]
\label{obs:resa_generalization}
If
\(
\mathbf{H}_k^{\mathrm{pre}}
=
\mathbf{H}_k^{\mathrm{post}}
=
\mathbf{H}_k^{\mathrm{res}}
=
\mathbf{I},
\)
then ReSA reduces to
\begin{equation}
\mathbf{A}_k
=
\mathbf{Z}_k
+
\phi\!\left(
\mathbf{Z}_k\mathbf{W}_{\downarrow}
\right)
\mathbf{W}_{\uparrow},
\label{eq:resa_plain_special_case}
\end{equation}
which is a residual bottleneck update in stream space.
\end{observation}

Observation~\ref{obs:resa_generalization} shows that ReSA retains the residual form while generalizing its fixed identity shortcut and unstructured feature flow with learnable stream-wise pre-, post-, and residual mappings. Together, Observations~\ref{obs:shaf_extension}--\ref{obs:resa_generalization} provide algebraic interpretations of memory transport in SHAF and state refinement in ReSA without requiring additional optimization claims.

\section{Experimental Setup}

\begin{table}
  \caption{Dataset Description.}
  \label{tab:dataset}
\renewcommand\tabcolsep{6pt}
\renewcommand{\arraystretch}{0.7}
  \begin{tabular}{cccc}
    \toprule
    \multirow{2}{*}{Dataset}&\multirow{2}{*}{Users}&\multirow{2}{*}{Items}&\multirow{2}{*}{Interaction}\\
    &&&\\
    \midrule
    MicroLens-100K&100,000&19,738&719,405\\
    MicroLens-50K&50,000&19,099&339,511\\
    Scientific&12,076&20,314&81,711\\
  \bottomrule
\end{tabular}
\vspace{-0.15in}
\end{table}

\noindent \textbf{Datasets.}
We evaluate on three datasets as demonstrated in Table \ref{tab:dataset}: MicroLens-50K, MicroLens-100K, and Scientific. MicroLens-50K and MicroLens-100K are public user-level subsets of the MicroLens benchmark for content-driven micro-video recommendation~\cite{ni2025content}. Scientific data from the Amazon dataset is included to test on other domains following~\cite{fu2024iisan}. For all datasets, we construct chronological user interaction sequences and pair them with the corresponding item-side multimodal content following~\cite{fu2024iisan}. 

\noindent \textbf{Implementation Details.}
Our frozen backbone family is Qwen3-VL-Embedding in two scales, 2B and 8B~\cite{li2026qwen3}. We strictly follow the experiment setup  in ~\cite{fu2024iisan}. Unless otherwise stated, the backbone is kept frozen, and only cached hidden states are used following~\cite{fu2024iisan}. Under this setting, only two key components are trained: the side adapter and the downstream sequential model. The downstream sequential recommender follows~\cite{fu2024iisan,hou2022towards,yuan2023go} and uses two Transformer blocks with two attention heads. Following ~\cite{fu2025crossan, fu2025efficient}, we optimize with AdamW with a weight decay of 0.1. We search the learning rate for the adapters from \{$1\times10^{-5}$, $1\times10^{-4}$,$1\times10^{-3}$\} and set to $1\times10^{-4}$, a batch size of 512. The bottleneck adapter rank was selected from \{128, 256, 512, 1024\}. User sequences are truncated to length 10. 

\noindent \textbf{Evaluation Protocol.}
We follow the standard leave-one-out evaluation protocol used in prior sequential recommendation work~\cite{kang2018self,ni2025content,fu2024iisan,fu2025crossan}. For each user, the last interaction is used for testing and the previous one for validation. We report ranking quality with HR@10, NDCG@10, HR@20, and NDCG@20~\cite{ni2025content}. All methods share the same data splits, downstream sequential model configuration, and full-ranking evaluation over the entire item set.

\section{Experimental results}
In this section, we report our experimental results and address the following research questions:
\begin{itemize}
    \item \textbf{RQ1:} Is Stresa better than a standard side adapter?
    \item \textbf{RQ2:} How does Stresa compare with state-of-the-art Side adaptation-based multimodal sequential recommendation baselines?
    \item \textbf{RQ3:} Which modules contribute to the gains?
    \item \textbf{RQ4:} What learning mechanism underlies the difference between Stresa and Side Adapter?
\end{itemize}

\subsection{Comparison with Side Adapters (RQ1)}

\begin{table}[t]
  \centering
  \caption{\textbf{Main results under frozen-backbone adaptation.} Two-Stage uses only the frozen final-layer backbone embedding. Side Adapter denotes the conventional decoupled side-branch baseline. Stresa consistently improves both regimes across Qwen3-VL-2B and Qwen3-VL-8B backbones and MicroLens-50K/100K. Best results in each backbone block are bold, and second-best results are underlined.  $^{*}$ indicates that Stresa significantly outperforms the second-best method at the 0.05 level based on a t-test over five random seeds}
  \vspace{-0.15in}
  \label{tab:main_results}
  \renewcommand{\arraystretch}{0.8}
  \setlength{\tabcolsep}{2pt}
  \begin{tabular}{@{}llcccc@{}}
    \toprule
    \textbf{Backbone} & \textbf{Method} & \textbf{H@10} & \textbf{N@10} & \textbf{H@20} & \textbf{N@20} \\
    \midrule
    \multicolumn{6}{c}{\emph{MicroLens-50K}} \\
    \midrule
    \multirow{3}{*}{\shortstack{Qwen3-VL-\\Embedding-8B}}
      & Two-Stage    & 0.0618 & 0.0335 & 0.0897 & 0.0405 \\
      & Side Adapter & \underline{0.0813} & \underline{0.0437} & \underline{0.1169} & \underline{0.0527} \\
      & \textbf{Stresa (Ours)}       & \textbf{0.0841}\textsuperscript{*} & \textbf{0.0456}\textsuperscript{*} & \textbf{0.1212}\textsuperscript{*} & \textbf{0.0549}\textsuperscript{*} \\
    \midrule
    \multirow{3}{*}{\shortstack{Qwen3-VL-\\Embedding-2B}}
      & Two-Stage    & 0.0595 & 0.0323 & 0.0878 & 0.0394 \\
      & Side Adapter & \underline{0.0728} & \underline{0.0397} & \underline{0.1042} & \underline{0.0476} \\
      & \textbf{Stresa (Ours)}      & \textbf{0.0783}\textsuperscript{*} & \textbf{0.0420}\textsuperscript{*} & \textbf{0.1145}\textsuperscript{*} & \textbf{0.0511}\textsuperscript{*} \\
    \midrule
    \multicolumn{6}{c}{\emph{MicroLens-100K}} \\
    \midrule
    \multirow{3}{*}{\shortstack{Qwen3-VL-\\Embedding-8B}}
      & Two-Stage    & 0.0706 & 0.0385 & 0.1006 & 0.0461 \\
      & Side Adapter & \underline{0.0977} & \underline{0.0531} & \underline{0.1398} & \underline{0.0637} \\
      & \textbf{Stresa (Ours)}       & \textbf{0.1019}\textsuperscript{*} & \textbf{0.0559}\textsuperscript{*} & \textbf{0.1442}\textsuperscript{*} & \textbf{0.0666}\textsuperscript{*} \\
    \midrule
    \multirow{3}{*}{\shortstack{Qwen3-VL-\\Embedding-2B}}
      & Two-Stage    & 0.0665 & 0.0365 & 0.0981 & 0.0444 \\
      & Side Adapter & \underline{0.0895} & \underline{0.0489} & \underline{0.1296} & \underline{0.0590} \\
      & \textbf{Stresa (Ours) }     & \textbf{0.0974}\textsuperscript{*} & \textbf{0.0541}\textsuperscript{*} & \textbf{0.1384}\textsuperscript{*} & \textbf{0.0644}\textsuperscript{*} \\
    \bottomrule
  \end{tabular}
    \vspace{-0.15in}

\end{table}

\begin{figure}
    \centering
    \includegraphics[width=\linewidth]{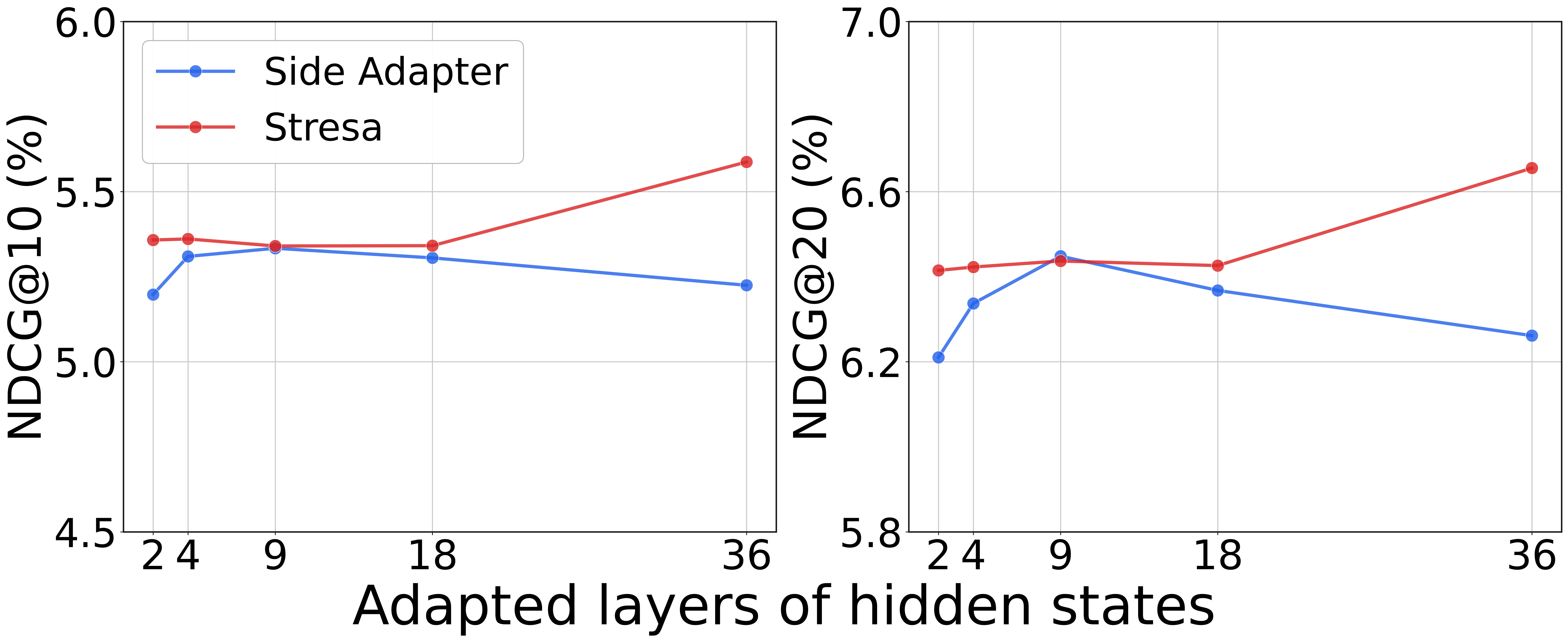}
    \vspace{-0.25in}
    \caption{NDCG@10 and NDCG@20 on MicroLens-100K (Qwen3-VL-8B) vs.\ the number of adapted layers of hidden states, for Side Adapter and Stresa. Metrics are taken at the epoch with the best validation HR@10.}
    \vspace{-0.2in}
    \label{fig:hidden_layer_adapt}
\end{figure}

\begin{table*}[t]
  \centering
  \caption{\textbf{Main results on three datasets.} Best results in each dataset are bold, and second-best results are underlined. $^{*}$ indicates that Stresa significantly outperforms the second-best method according to a $t$-test ($p<0.05$) over five seeds.}
  \vspace{-0.15in}
  \label{tab:main_results_three}
  \setlength{\tabcolsep}{2pt}
\renewcommand{\arraystretch}{0.8}
  \begin{tabular}{@{}lcccccccccccc@{}}
    \toprule
    \multirow{2}{*}{\textbf{Method}} 
    & \multicolumn{4}{c}{\textbf{MicroLens-50K}} 
    & \multicolumn{4}{c}{\textbf{MicroLens-100K}} 
    & \multicolumn{4}{c}{\textbf{Scientific}} \\
    \cmidrule(lr){2-5} \cmidrule(lr){6-9} \cmidrule(lr){10-13}
    & \textbf{H@10} & \textbf{N@10} & \textbf{H@20} & \textbf{N@20}
    & \textbf{H@10} & \textbf{N@10} & \textbf{H@20} & \textbf{N@20}
    & \textbf{H@10} & \textbf{N@10} & \textbf{H@20} & \textbf{N@20} \\
    \midrule
    IISAN~\cite{fu2024iisan} (SIGIR'24)
      & 0.0771 & 0.0421 & 0.1121 & 0.0509
      & 0.0960 & 0.0526 & 0.1366 & 0.0628
      & 0.0683 & 0.0414 & 0.0933 & 0.0470 \\
    XSMoE~\cite{qu2025efficient} (CIKM'25)
      & 0.0550 & 0.0296 & 0.0807 & 0.0361
      & 0.0697 & 0.0376 & 0.1037 & 0.0461
      & 0.0620 & 0.0395 & 0.0840 & 0.0450 \\
    CROSSAN~\cite{fu2025crossan} (arXiv'25)
      & \underline{0.0806} & \underline{0.0431} & \underline{0.1177} & \underline{0.0524}
      & \underline{0.0999} & \underline{0.0553} & \underline{0.1428} & \underline{0.0661}
      & 0.0690 & 0.0422 & 0.0928 & 0.0482 \\
    IISAN-Versa~\cite{fu2025efficient} (TKDE'25)
      & 0.0784 & 0.0421 & 0.1126 & 0.0507 
      & 0.0964 & 0.0533 & 0.1341 & 0.0618 
      & \underline{0.0739} & \underline{0.0447} & \underline{0.1014} & \underline{0.0516} \\
    \textbf{Stresa (Ours)}
      & \textbf{0.0841}\textsuperscript{*} & \textbf{0.0456}\textsuperscript{*} & \textbf{0.1212}\textsuperscript{*} & \textbf{0.0549}\textsuperscript{*}
      & \textbf{0.1019}\textsuperscript{*} & \textbf{0.0559}\textsuperscript{*} & \textbf{0.1442}\textsuperscript{*} & \textbf{0.0666}\textsuperscript{*}
      & \textbf{0.0794}\textsuperscript{*} & \textbf{0.0483}\textsuperscript{*} & \textbf{0.1067}\textsuperscript{*} & \textbf{0.0551}\textsuperscript{*} \\
    \bottomrule
  \end{tabular}
  \vspace{-0.15in}
\end{table*}
RQ1 asks whether Stresa is better than a standard Side Adapter. To answer this question, we compare Stresa against both a conventional Side Adapter~\cite{fu2024iisan,fu2025crossan,fu2025efficient} and a Two-Stage baseline~\cite{yuan2023go} that uses only the frozen final-layer backbone embedding. The comparison is conducted on both MicroLens-50K and MicroLens-100K, with two backbone scales, Qwen3-VL-2B and Qwen3-VL-8B, and we report the best test performance under each setting. In addition, to directly verify our motivation of adapting more of the backbone's potential hidden representations, we further vary the number of adapted hidden-state layers and compare the resulting trends between Side Adapter and Stresa in Fig.~\ref{fig:hidden_layer_adapt}.

The results in Table~\ref{tab:main_results} show that Stresa consistently outperforms both Two-Stage and the standard Side Adapter across all backbone and dataset settings. The gains over Two-Stage confirm the value of adaptation beyond frozen embeddings, while the gains over Side Adapter show that the improvement comes from a better adaptation design rather than simply using side information. More importantly, Fig.~\ref{fig:hidden_layer_adapt} reveals a clear difference in depth behavior. Stresa remains stable as adaptation extends to more hidden layers and achieves the best performance when adapting all layers, while its degradation is still limited even when only a small number of layers are adapted. In contrast, the standard Side Adapter reaches its best performance around 9 adapted layers, and then declines noticeably as more layers are included. This supports our claim that standard side adaptation does not fully exploit deeper hidden states, whereas Stresa can better utilize the full adaptation space while maintaining stability.

\textbf{Answer to RQ1:} Yes. Stresa consistently outperforms the standard Side Adapter across datasets and backbone scales. Its advantage is particularly evident in deeper adaptation settings, where it can stably exploit more hidden states and better unlock the full potential of the backbone.

\begin{table}[t]
\centering
\caption{\textbf{Ablation on Qwen3-VL-8B.} The progression from a standard side adapter to Stresa.}
\vspace{-0.15in}
\setlength{\tabcolsep}{1pt}
\renewcommand{\arraystretch}{0.8}
\label{tab:ablation}
\begin{tabular}{lcccc}
\toprule
Method & HR@10 & NDCG@10 & HR@20 & NDCG@20 \\
\midrule
\multicolumn{5}{c}{\emph{MicroLens-50K}} \\
\midrule
 Side Adapter & 0.0813 & 0.0437 & 0.1169 & 0.0527 \\
+ ReSA & 0.0829 & 0.0444 & 0.1188 & 0.0534 \\
+ ReSA + SHAF (Stresa) & \textbf{0.0841} & \textbf{0.0456} & \textbf{0.1212} & \textbf{0.0549} \\
\midrule
\multicolumn{5}{c}{\emph{MicroLens-100K}} \\
\midrule
 Side Adapter & 0.0977 & 0.0531 & 0.1398 & 0.0637 \\
+ ReSA & 0.0999 & 0.0545 & 0.1417 & 0.0650 \\
+ ReSA + SHAF (Stresa)  & \textbf{0.1019} & \textbf{0.0559} & \textbf{0.1442} & \textbf{0.0666} \\
\bottomrule
\end{tabular}
\vspace{-0.15in}
\end{table}

\begin{table}[t]
\centering
\caption{Ablation on Identity Preservation.}
\vspace{-0.15in}
\setlength{\tabcolsep}{1pt}
\renewcommand{\arraystretch}{0.8}
\begin{tabular}{lcccc}
\toprule
Method & HR@10 & NDCG@10 & HR@20 & NDCG@20 \\
\midrule
\multicolumn{5}{c}{\emph{MicroLens-50K}} \\
\midrule
w/ Identity Preservation  & \textbf{0.0841} & \textbf{0.0456} & \textbf{0.1212} & \textbf{0.0549} \\
w/o Identity Preservation & 0.0807 & 0.0433 & 0.1184 & 0.0528 \\
\midrule
\multicolumn{5}{c}{\emph{MicroLens-100K}} \\
\midrule
w/ Identity Preservation  & \textbf{0.1019} & \textbf{0.0559} & \textbf{0.1442} & \textbf{0.0666} \\
w/o Identity Preservation & 0.0991 & 0.0545 & 0.1416 & 0.0652 \\
\bottomrule
\end{tabular}
\label{tab:identity_preservation}
\vspace{-0.1in}
\end{table}

\subsection{Comparison with SOTA baselines (RQ2)}

RQ2 asks how Stresa compares with state-of-the-art side adaptation-based multimodal sequential recommendation baselines. To this end, we compare our approach with IISAN~\cite{fu2024iisan}, XSMoE~\cite{qu2025efficient}, IISAN-Versa~\cite{fu2025efficient}, and CROSSAN~\cite{fu2025crossan}.\footnote{We focus this comparison on the side adaptation paradigm because our goal is to study a fully ID-free setting, where recommendation is driven only by multimodal side information without introducing additional ID signals or auxiliary supervision. Many existing multimodal sequential recommendation methods incorporate extra information beyond this setting, making direct comparison less fair. Moreover, side adaptation itself has already shown strong performance in prior work, making it a strong and meaningful basis for comparison. Therefore, for fairness and for a cleaner evaluation of the target setting, we mainly compare against representative side adaptation-based methods.} In these comparisons, Stresa uses Qwen3-VL-8B as the backbone. For the competing baselines, we preserve their original architectural settings and report their best results under those settings, because these methods were proposed in multi-tower side-adaptation frameworks, whereas our method is built on a single-tower Qwen3-VL-Embedding backbone. To provide a more direct reference under the same single-tower setting, we also include the Side Adapter baseline in the previous section. For IISAN-Versa, which considers the asymmetric case but still follows a two-tower design, we further evaluate a variant with Qwen3-VL-Embedding-8B on the text side to improve comparability.

Table~\ref{tab:main_results_three} shows that Stresa consistently achieves the best overall performance across all datasets. On MicroLens-50K and MicroLens-100K, it slightly but consistently surpasses the strongest baseline, CROSSAN, on all four metrics. On Scientific, Stresa improves HR@10 from 0.0739 to 0.0794 and NDCG@20 from 0.0516 to 0.0551 over the strongest prior baseline. Overall, these gains are modest but stable across datasets, showing that Stresa generalizes well across different scales and domains.

\textbf{Answer to RQ2:} Stresa consistently outperforms strong side adaptation-based baselines. This shows that its gains persist not only over a standard Side Adapter, but also over stronger side-adaptation variants, supporting the effectiveness of Stresa across competitive comparison settings.

\begin{table}[t]
\centering
\caption{Ablation on stream-aware residual.}
\vspace{-0.1in}
\setlength{\tabcolsep}{1pt}
\renewcommand{\arraystretch}{0.8}
\begin{tabular}{lcccc}
 \toprule
    Method & HR@10 & NDCG@10 & HR@20 & NDCG@20 \\
\midrule
\multicolumn{5}{c}{\emph{MicroLens-50K}} \\
\midrule
w/o Manifold Constraint & \textbf{0.0829} & 0.0444 & \textbf{0.1188} & 0.0534 \\
w/ Manifold Constraint & 0.0826 & \textbf{0.0456} & 0.1186 &\textbf{0.0547} \\
\midrule
\multicolumn{5}{c}{\emph{MicroLens-100K}} \\
\midrule
w/o Manifold Constraint & \textbf{0.0999} & 0.0545 & 0.1417 & 0.0650 \\
w/ Manifold Constraint & 0.0996 & \textbf{0.0546} & \textbf{0.1424} & \textbf{0.0655} \\
\bottomrule
    \end{tabular}
    \label{tab:resa_manifold}
    \vspace{-0.1in}
\end{table}

\subsection{Ablation Study (RQ3)}

RQ3 investigates whether Stresa is driven by a single dominant component or by the complementary effects of its modules. To answer this question, we perform comprehensive ablation studies in Tables~\ref{tab:ablation}, and \ref{tab:identity_preservation}, examining the contributions of ReSA, SHAF, and additional design choices, including the manifold-constrained residual connection~\cite{xie2025mhc} in ReSA and identity preservation in SHAF.

The results show a clear progression. In Table~\ref{tab:ablation}, adding ReSA on top of the Side Adapter consistently improves performance on both datasets, and further adding SHAF (Stresa) yields the best overall results across all metrics. For example, on MicroLens-100K, HR@10/NDCG@10 improve from 0.0977/0.0531 with Side Adapter to 0.0999/0.0545 with ReSA, and further to 0.1019/0.0559 with SHAF; similar gains are observed on HR@20/NDCG@20 and on MicroLens-50K. In Table~\ref{tab:resa_manifold} further evaluates a ReSA variant with the recently proposed manifold-constrained residual connection~\cite{xie2025mhc}. Although it brings modest gains on several NDCG metrics, the overall performance is comparable while incurring higher computational cost due to manifold operations. We thus do not use it as the default design. Table~\ref{tab:identity_preservation} shows that identity preservation is critical, and removing it causes substantial performance drops.

\textbf{Answer to RQ3:} The gains come from complementary modules: both ReSA and SHAF are important for achieving them.

\subsection{Visualization (RQ4)}
\label{sec:visualization}
\begin{figure}
    \centering
    \includegraphics[width=\linewidth]{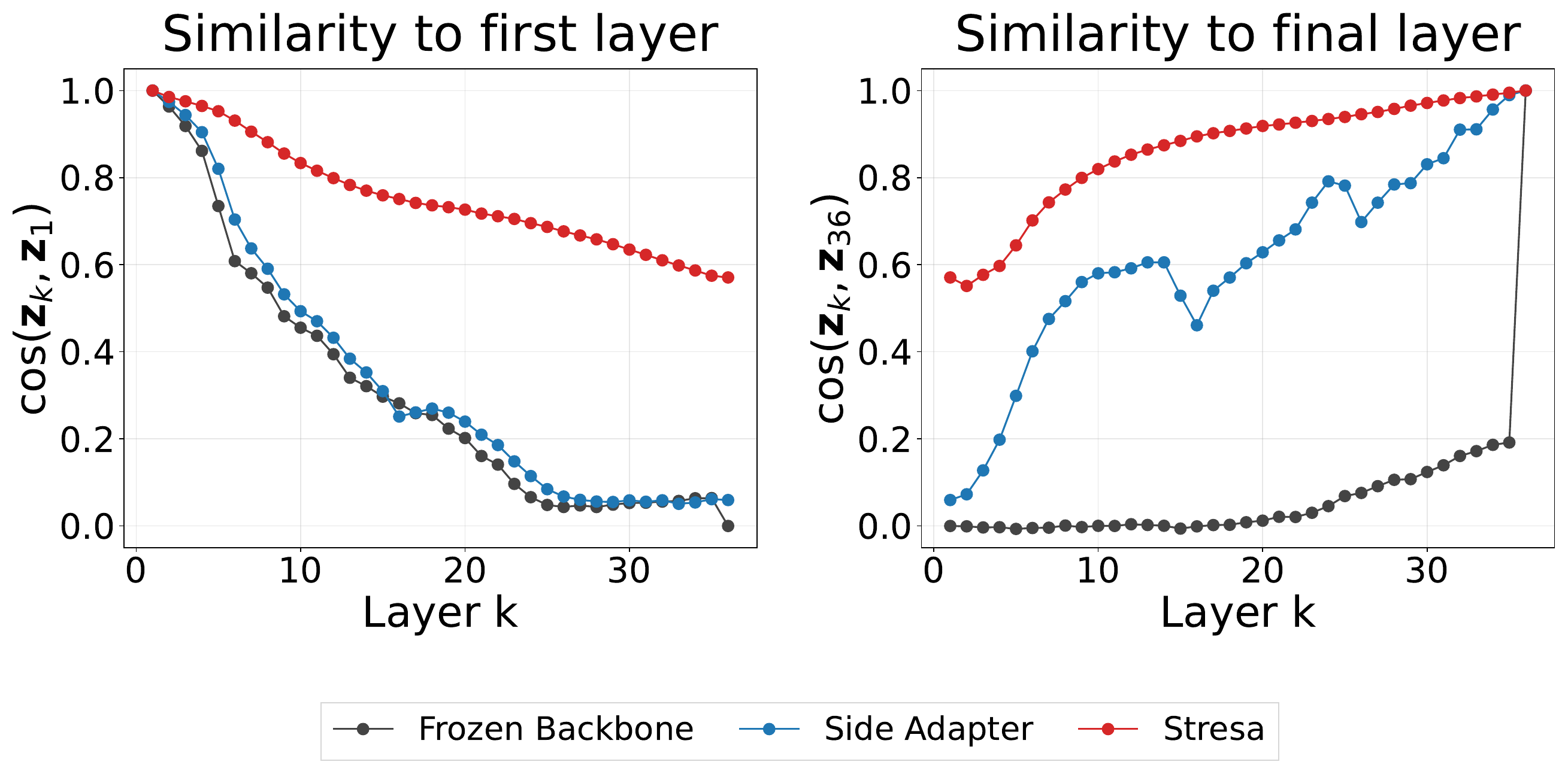}
    \vspace{-0.15in}
    \caption{Layer-wise cosine similarity on MicroLens-100K for the frozen Qwen3-VL backbone, Side Adapter, and Stresa. Left: $\cos(\mathbf{z}_k,\mathbf{z}_1)$ versus backbone layer index $k$. Right: $\cos(\mathbf{z}_k,\mathbf{z}_{36})$ versus $k$. Each series averages over items; layers $k=1,\ldots,36$ correspond to the 36 adapter depths.}
    \label{fig:curves}
    \vspace{-0.15in}
\end{figure}

To answer this question, we analyze layer-wise representation similarity for the frozen backbone, Side Adapter, and Stresa by comparing each layer’s cosine similarity to the first and final layers.
 Figure~\ref{fig:curves} shows that Stresa changes representations more smoothly across depth. Compared to Side Adapter, its similarity to the first layer decreases less abruptly, indicating better preservation of early information, while its similarity to the final layer increases more progressively, indicating stable layer-wise refinement rather than sharp late-stage shifts. 

\textbf{Answer to RQ4:} The underlying learning mechanism is that Stresa better reconciles identity preservation with progressive adaptation. This leads to a more stable representation trajectory across layers and, consequently, stronger adaptation performance, especially for larger backbones.

\begin{table}[!htbp]
\centering
\caption{Efficiency comparison between Stresa and Side Adapter.}
\vspace{-0.15in}
\renewcommand{\arraystretch}{0.8}
\label{tab:efficiency}
\begin{tabular}{lccc}
\toprule
Method & Params & Time (s/epoch) & FLOPs/step ($10^{12}$) \\
\midrule
Stresa & 78.06M & 392.41 & 5.311 \\
Side Adapter & 155.73M & 338.20 & 4.395 \\
\bottomrule
\end{tabular}
\vspace{-0.15in}
\end{table}

\section{Efficiency}

Table~\ref{tab:efficiency} reports the efficiency comparison between Stresa and the standard Side Adapter. Stresa reduces the number of trainable parameters, while introducing a modest increase in computation due to the additional stream-aware routing and mixing operations. In practice, this overhead is not significant: the increases in training time per epoch and FLOPs per step remain moderate, while Stresa achieves stronger recommendation performance in the main experiments.

\section{Conclusion and Future Work}

In this paper, we study how to make side adaptation for large pre-trained multimodal embedding models more effective under deep adaptation. We propose \emph{Stresa}, a stream-aware side-adaptation framework achieving selective residual update and earlier represtation preservation fusion. Across our experiments, Stresa consistently outperforms frozen two-stage adaptation, standard Side Adapter, and strong side adaptation-based baselines. The results further show that its advantage is especially clear when adaptation goes deeper, suggesting that stable memory preservation and stream-aware refinement are key to fully exploiting hidden states in frozen multimodal backbones. Overall, these findings indicate that stream-aware side adaptation is a promising direction for multimodal sequential recommendation.

For future work, it would be valuable to evaluate Stresa on broader multimodal backbones, datasets, and recommendation domains, and to explore more adaptive strategies for selecting or scheduling adapted layers. Another important direction is to connect this framework with more practical deployment settings, such as online or continual adaptation. We hope this work can encourage further research on stable and efficient adaptation for frozen multimodal foundation models in recommendation.
\balance
\bibliographystyle{ACM-Reference-Format}
\bibliography{sample-base}

\end{document}